\documentclass[twocolumn,english]{article}
\usepackage[T1]{fontenc}
\usepackage{amssymb}
\usepackage{amsmath}
\usepackage{graphicx}
\usepackage{babel}

\makeatletter
\usepackage{a4}
\input{epsfig.sty}
\def\fnum@table{\tablename~{\bf\thetable}}
\def\fnum@figure{\figurename~{\bf\thefigure}}
\def\tablename{\footnotesize{\bf Table}}
\def\figurename{\footnotesize{\bf Figure}}

\def\be{\begin{equation}}
\def\ee{\end{equation}}

\makeatother

\usepackage{babel}
\begin{document}

\title{\textbf{QGSJET-III model of high energy hadronic interactions: II. Particle
production and extensive air shower characteristics}}

\author{Sergey Ostapchenko\\
\textit{\small Universit\"at Hamburg, II Institut f\"ur Theoretische
Physik, 22761 Hamburg, Germany}\\
}

\maketitle
\textbf{Abstract}
The hadronization procedure of the QGSJET-III Monte Carlo (MC) generator 
 of high energy hadronic interactions is discussed.
 Selected results of the model, regarding production spectra of 
 secondary particles, are presented in comparison to experimental data
 and to the corresponding predictions of the QGSJET-II-04 MC generator.
 The model is applied to calculations of basic characteristics of 
 extensive air showers initiated by cosmic ray interactions in the 
 atmosphere and the results are compared to predictions of other
 MC generators of cosmic ray interactions.

\section{Introduction\label{intro.sec}}
All current Monte Carlo (MC) generators of high energy
hadronic interactions rely on the qualitative picture of quantum chromodynamics (QCD) in the sense that such
interactions are assumed to be mediated by parton cascades developing 
between the colliding
  hadrons (nuclei). Particle production occurs when the coherence of (some of)
  these parton cascades is broken by the scattering process and final $s$-channel
  partons convert into hadrons.  Since such a hadronization 
  is essentially a nonperturbative process,  the corresponding  treatment
   relies on phenomenological approaches. The most popular ones are   string
   fragmentation procedures implying a formation of strings of color
   field between  final partons. With such partons flying apart, the 
   tension of a string rises until it breaks up, with the color 
   field being neutralized
   via a creation of quark-antiquark and diquark-antidiquark pairs from the
   vacuum, ending up with a formation of colorless parton clusters identified
   with hadrons  \cite{and83}.

Needless to say, modeling such a process involves a substantial number of 
adjustable parameters and requires a thorough calibration with various
experimental data. Nevertheless, this does not necessarily spoil the predictive
power of  hadronic interaction models since the mechanism is
 assumed to be universal. E.g., moving
to higher energies or replacing the interacting hadrons, say, protons, by
other hadrons or by nuclei, one arrives to  a different parton configuration.
However, the hadronization procedure remains unaffected by such changes, hence,
the conversion of partons into hadrons can be described using one and the same
approach, regardless the type and kinematics of the interaction. This allows 
one to test both the validity of an interaction model itself and the quality
of the hadronization mechanism employed and, in case both perform satisfactorily
 enough, to use the complete model for predictive calculations, e.g., regarding 
the properties of  so-called   extensive air
showers (EAS) initiated by interactions of 
high energy cosmic rays in the atmosphere (see  
  \cite{eng11} for a review).

In the particular case of the QGSJET-III MC generator  \cite{ost24}, 
the treatment of the
hadronization follows closely the one of the original QGSJET model \cite{kal93,kal97}, with a number of technical improvements, as discussed in Section \ref{hadr.sec}. In Section \ref{results.sec}, a comparison with selected
accelerator data will be presented, followed  in Section \ref{pn.sec}
by a discussion of basic properties of proton-nitrogen interactions,
relevant for EAS predictions. In turn, Section  \ref{eas.sec} is devoted to
calculations of extensive air shower characteristics.
The conclusions will be given in Section  \ref{concl.sec}.

\section{Hadronization procedure of the QGSJET-III model\label{hadr.sec}}
As described in \cite{ost24}, the procedure for simulating a particular hadronic
(nuclear) collision event in the  QGSJET-III model consists of two parts.
First, a ``macro-configuration'' for an inelastic interaction is defined, i.e.,
the complete ``net'' of ``elementary'' 
parton cascades giving rise to secondary hadron production and being
 described by cut Pomerons is reconstructed, based on the corresponding partial cross sections. The $t$- and $s$-channel evolution of perturbative
  parton cascades corresponding to parton
virtualities $|q^2|>Q_0^2$, $Q_0^2$ being the ``infrared'' cutoff of the model,
is treated explicitly.\footnote{The modeling of the final (s-channel)
parton emission follows closely the procedure 
described in \cite{mar84}, imposing the final transverse 
momentum cutoff $p^2_{t,{\rm cut(f)}} = 0.15$ GeV$^2$.}
At the second step, one considers a formation of two
strings of color field per cut Pomeron, stretched between constituent partons
of the interacting hadrons (nuclei) or/and between the final $s$-channel partons
resulting from the treatment of perturbative parton cascades.
The breakup and hadronization of those strings are modeled
by means of a string fragmentation procedure.

While the 4-momenta of final  $s$-channel partons resulting from  perturbative
 parton cascades are already defined at the first step, this is not the case
for constituent partons. For a given projectile or target hadron $h$, to which
$2n$ strings are attached, the sharing of the, respectively, light cone plus 
(LC$^+$) or  light cone minus (LC$^-$) momentum fractions $x^{\pm}$ between
constituent partons (string ``ends'') is performed according to the
 distribution
 \begin{eqnarray}
 f_h^{\rm LC}(x_1^{\pm},...,x_{2n}^{\pm})\propto \left[\prod_{i=1}^{2n-2}
 (x_i^{\pm})^{-\alpha_{\rm sea}}  \right] \nonumber \\
 \times \; (x_{2n-1}^{\pm})^{-\alpha_{\mathbb{R}}}\,
 (x_{2n}^{\pm})^{-\alpha_{\rm r}}\, \delta \!\left(1-\sum_{j=1}^{2n}x_j^{\pm}\right) .
 \label{ems.eq}
 \end{eqnarray}
 Here the exponent $\alpha_{\rm sea}$ is used for sea (anti)quarks, considering
 only  light $u$ and $d$ (anti)quarks as string ends, while the small $x$ limit
 of a valence (anti)quark distribution  follows the Regge behavior 
 ($\propto x_{2n-1}^{-\alpha_{\mathbb{R}}}$), 
 with $\alpha_{\mathbb{R}}=\alpha_{\rho}(0)=0.5$.
 Similarly, the LC momentum distribution of the remaining valence parton
 [(anti)quark for $h$ being a meson or (anti)diquark for a baryon] follows the
 corresponding Regge behavior ($\propto x_{2n}^{-\alpha_{\rm r}}$) \cite{kai82,kai86,kai87}:
  \begin{eqnarray}
 \alpha_u  =\alpha_d  = \alpha_{\mathbb{R}} 
  \label{alpha-u.eq}\\
  \alpha_{ud}  =\alpha_{uu}+1  =2\alpha_{N}(0)- \alpha_{\mathbb{R}} \\ 
  \alpha_s  = \alpha_{\phi}(0)\\
  \alpha_{us}  =\alpha_{ds}  = \alpha_{ud}-\alpha_{\mathbb{R}} +  \alpha_{\phi}(0) \,,
 \label{alpha-us.eq}
   \end{eqnarray}
  the corresponding Regge intercepts being compiled in Table~\ref{Flo:regge-interc}.
     \begin{table}[htb]
\begin{centering}
\begin{tabular}{|ccccc|}
\hline 
$\alpha_{\rho}(0)$ &   $\alpha_{\phi}(0)$ &  
  $\alpha_{N}(0)$ &  $\alpha_{\pi}(0)$ &  $\alpha_{f}(0)$
   \tabularnewline
\hline 
  0.5 & 0 &  0 & 0 &  0.7
\tabularnewline
\hline
\end{tabular}\caption{Intercepts of selected  Regge trajectories.}
\label{Flo:regge-interc}
\par\end{centering}
 \end{table}

Regarding the   LC momentum distribution of constituent sea quarks,
in principle, it should also be characterized by the Regge behavior
 ($\propto x_i^{-\alpha_{\mathbb{R}}}$) in the  low $x$ limit \cite{kai82},
 which was adopted in the QGSJET \cite{kal93,kal97} and QGSJET-II \cite{ost06,ost11}
 models. However, here we treat the respective exponent as an adjustable parameter,  
  $0.5 \leq \alpha_{\rm sea}<1$. The reason is that we consider
 Pomeron-Pomeron interactions, introducing a minimal Pomeron rapidity ``size''
 $\xi$ \cite{ost24}, for the Pomeron asymptotics to be applicable. Therefore, the
 deviation of $\alpha_{\rm sea}$  from $\alpha_{\mathbb{R}}$ is assumed to
 take effectively into account a formation of small rapidity gaps
  $y_{\rm gap}<\xi$, not treated explicitly by the model. 
  
 In turn,  transverse momenta of string ends are sampled according to the distribution:
   \begin{eqnarray}
 f_h^{\perp}(\vec p_{t_1},...,\vec p_{t_{2n}})
 \propto 
 \exp\!\left(-\sum_{i=1}^{2n-1}\frac{p_{t_i}^2}{\gamma_{\perp}}
 -\frac{p_{t_{2n}}^2}{\gamma_{\perp}^h}\right) \nonumber \\
\times \; \delta^{(2)}\!\left(1-\sum_{j=1}^{2n}\vec p_{t_j}\right),
\label{pts.eq}
 \end{eqnarray}
the corresponding parameters being listed in Table~\ref{hadr.tab}.
 
  \begin{table*}[t]
\begin{centering}
\begin{tabular}{|cccccccccccc|}
\hline 
   $\alpha_{\rm sea}$  &   $w^{\pi}_p$   &  $w^{\pi}_{\pi}$ &   $w_{K}^{\pi}$ 
  &   $w^{f}_p$   &   $w^{f}_{\pi}$  &   $w^{f}_{K}$  &   $\gamma_{\perp}$
   &  $\gamma_{\perp}^p$   &   $\gamma_{\perp}^{\pi}$  &  $\gamma_{\perp}^{K}$ 
  &   $p^2_{t,{\rm cut(f)}}$  
  \tabularnewline
& & & & &     &
 &  {\scriptsize GeV$^2$}& {\scriptsize GeV$^2$} & {\scriptsize GeV$^2$} &  {\scriptsize GeV$^2$} &  {\scriptsize GeV$^2$}
  \tabularnewline
\hline 
  0.65  &   0.28  &  0.37  &   0.11  &    0.08  &   0.15  &  0  
  &   0.15  &   0.72  &  0.2  &   0.2   &  0.15 
\tabularnewline
\hline
\end{tabular}\caption{Parameters of the energy-momentum sharing and   hadronization  procedures of the QGSJET-III model
(excluding the parameters of the string fragmentation procedure).}
\label{hadr.tab}
\par\end{centering}
 \end{table*}

 \begin{table*}[t]
\begin{centering}
\begin{tabular}{|cccccccccc|}
\hline 
  $\Lambda$  &   $p^2_{t,{\rm thr}}$   &  $a_{ud}$  &    $a_{us}=a_{ds}$  &  $a_{s}$  &  
   $\gamma_{ud}=\gamma_{us}=\gamma_{ds}$
  &  $\gamma_{u}=\gamma_{d}$  &   $\gamma_{s}$    &   $w_{\eta}$    &   $w_{\rho}$ 
  \tabularnewline
  &  {\scriptsize GeV$^2$}  && && {\scriptsize GeV} &  {\scriptsize GeV} &  {\scriptsize GeV}  
 &&
  \tabularnewline
\hline 
  1.55 &   0.55  &   0.085  &  0.038  &   0.125  &   0.33  &    0.21  &   0.28  &   0.11  &   0.33 
\tabularnewline
\hline
\end{tabular}
\caption{Parameters of the string fragmentation procedure of the QGSJET-III model.}
\label{string-hadr.tab}
\par\end{centering}
 \end{table*}

\begin{figure}[htb]
\centering
\includegraphics[height=2.4cm,width=0.49\textwidth]{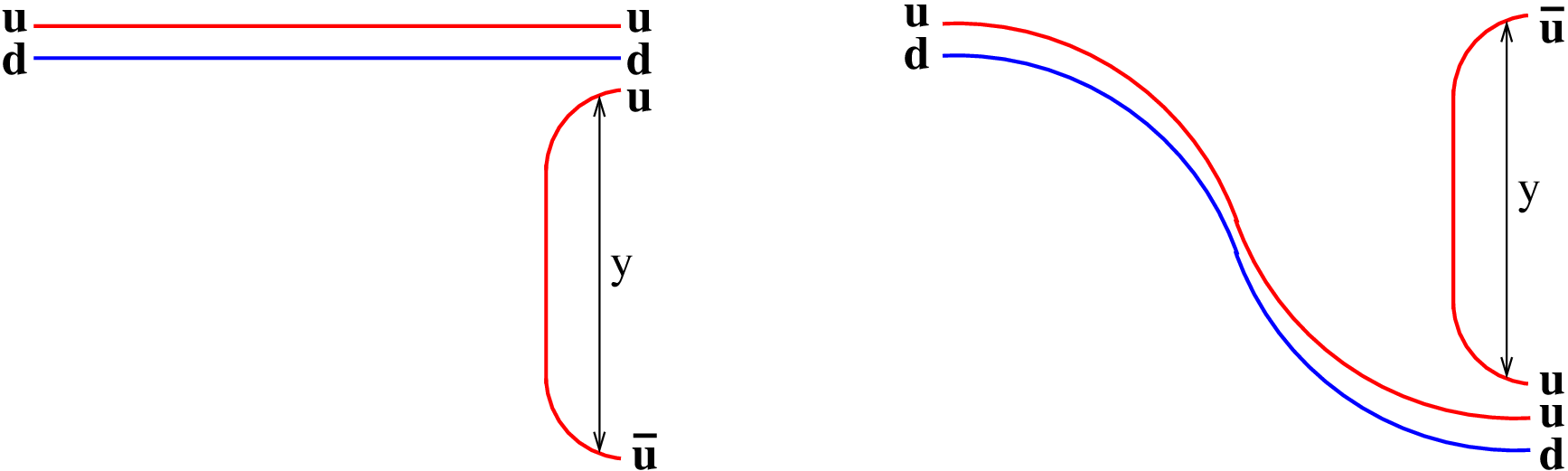}
\caption{Schematic view of the fragmentation of a fast $ud$ diquark into a proton:
both large $x$ and small $x$ limits of the proton LC momentum distribution
correspond to a large rapidity $y$ separation between the $u$ and $\bar u$
of the vacuum-created $u\bar u$ pair. Left: the large $x$ limit  
($1-x\propto e^{-y}$) is obtained  by slowing down the  $\bar u$ antiquark.
Right: the small  $x$ limit ($x\propto e^{-y}$) corresponds to slowing down 
both the $u$ quark and the original $ud$ diquark.}
\label{fig:fragm}       
\end{figure}

The fragmentation of a string into secondary hadrons is performed considering
iteratively an emission of a hadron from either string end (see, e.g.\ \cite{wer89}), 
with the hadron LC momentum fraction $x$ (LC$^+$ or LC$^-$ for a projectile or
 target side end, respectively) and the transverse momentum
 $p_t$ in the center-of-mass (c.m.) frame of the string being sampled 
 according to the distribution:
   \begin{eqnarray}
 f_{qq'\rightarrow h}^{\rm fragm}(x,\vec p_{t}) 
 \propto x^{1-\alpha_q-\alpha_{q'}}(1-x)^{\Lambda-\alpha_{q'}} 
 e^{-\frac{p_{t}}{\gamma_{q'}}}\!,
\label{fragm.eq}
 \end{eqnarray}
using Eqs.\ (\ref{alpha-u.eq}-\ref{alpha-us.eq}) for the relations between the
parameters  $\alpha_q$ and  the intercepts of the corresponding Regge
 trajectories. Here the large $x$ limit is defined by the probability to slow
 down (move away in rapidity space)
  the antiparton [(anti)quark or (anti)diquark] $\bar q'$ 
of the vacuum-created
 pair, while the small $x$ limit is obtained by slowing down both the original
 parton $q$ and the parton $q'$ attached to it to form the final hadron
 ($q+  q'\rightarrow h$) \cite{kai87} (cf.\ Fig.\ \ref{fig:fragm}). 
 Various parton-antiparton pairs $\bar q'q'$
 created from the vacuum are sampled according to the corresponding probabilities
  $a_{q'}$, relative to $\bar u u$ and $\bar dd$ pairs ($a_u=a_d=1/2$).
 The iterative string fragmentation is continued until
  the mass  of the string
 reminder falls below a specified threshold\footnote{
 We use $M_{\rm thr}^{qq'}=\sum _{i=1}^3 \sqrt{m_{h_i}^2+p^2_{t,{\rm thr}}}$,
where $h_1$,  $h_2$, and $h_3$ correspond to the minimal mass 3-hadron final
state  for the types $q$ and $q'$ of the string end  partons.}
 $M_{\rm thr}$,  followed by a modeling of
  a two particle decay of the remaining string. The string fragmentation
  parameters are compiled  in Table \ref{string-hadr.tab}. 
 
The above-discussed treatment takes into account the production of final
``stable'' hadrons: pions, kaons, nucleons, lambda and sigma baryons,
including the corresponding antiparticles, where relevant. 
Contributions of decays of short-lived resonances are assumed here to be
accounted for in the stable hadron yields via the duality principle. The
exceptions are the formation of $\Delta^{++}$ and  $\Delta^{-}$ resonances
(and the corresponding antiparticles)
by a fragmentation of constituent $uu$ and $dd$ (anti)diquarks as well as the
production of  $\rho$ and $\eta$ mesons, which is also treated explicitly.
 Regarding the latter, the probability to produce 
 an $\eta$ meson, relative to $\pi^0$,
is chosen as $w_{\eta}=1/9$ from the quark counting rules. Similarly, 
we use $w_{\rho}=1/3$ for the corresponding probability of $\rho$
mesons, relative to pions, 
assuming that $\sim 50$\% of final pions originate from $\rho$ decays.
   For brevity, we restrain here
 from discussing other technicalities of the hadronization procedure.

\begin{figure*}[t]
\centering
\includegraphics[height=6.cm,width=\textwidth]{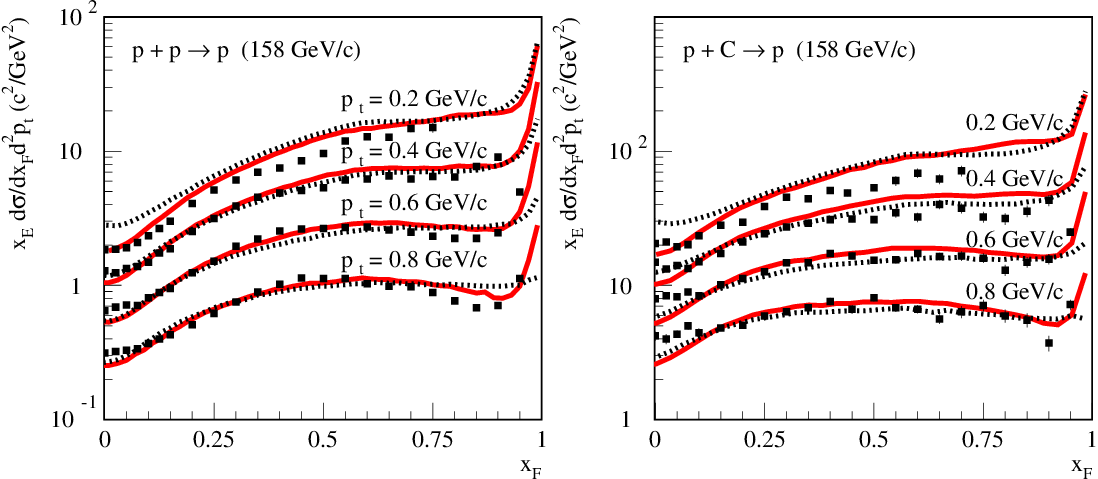}
\caption{$x_{\rm F}$-dependence of invariant cross sections in c.m.\ frame,
$x_E d\sigma /dx_{\rm F}/d^2p_t$ ($x_E=2E/\sqrt{s}$),
 for different $p_t$ values (as indicated in the plots), for proton production in
$pp$ (left) and $p$C (right) collisions at 158 GeV/c, as calculated using the 
QGSJET-III (solid lines) and QGSJET-II-04 (dotted lines) models, compared to 
NA49 data \cite{ant10,baa13} (points).}
\label{fig:prot158}       
\end{figure*}%

In addition to the basic procedure described above, a treatment of the pion
exchange contribution to forward hadron production, i.e.,
 the Reggeon-Reggeon-Pomeron ($\mathbb{RRP}$) process, with $\mathbb{R}=\pi$, has been
considered in the QGSJET-III model\footnote{A simplified treatment of the pion
exchange process for pion-proton and pion-nucleus collisions had earlier been implemented in the QGSJET-II-04 model \cite{ost13}.} \cite{ost21}. As discussed
in more detail in \cite{ost21}, due to the low intercept of the pion Regge
trajectory, $\alpha_{\pi}(0)\simeq 0$, the process does not lead to a formation
of a large rapidity gap between the ``leading'', i.e., most energetic, secondary
hadron and the remaining hadronic system, with the consequence that absorptive
corrections to the corresponding ``bare'' triple-Reggeon contribution are
essentially the same as for the single cut Pomeron process. This motivated us 
to sample the $\mathbb{RRP}$ process, for $\mathbb{R}=\pi$, as a fixed fraction,
with probability $w^{\pi}_h$, of the single cut Pomeron contribution, for a 
given hadron $h$. The hadronization procedure remains practically unaltered in
such a case, except that the leading hadron is sampled according to the 
respective $x$- and $p_t$-distribution for the $\pi$-exchange
process \cite{ost21}, while the
pair of strings is  attached here to the quark and antiquark of the virtual
pion, instead of constituent partons of the initial hadron $h$. In other words,
after the production of the leading hadron, the rest of the final state
corresponds to an inelastic interaction\footnote{Since  absorptive 
corrections ``push'' the $\pi$-exchange process towards large impact parameters,
the contribution of an elastic scattering of the virtual pion is strongly
suppressed and can be neglected  \cite{ost21}.} 
of the virtual pion with the target (in case $h$ is a projectile
 hadron), with the respectively reduced c.m.\ energy for the 
collision. Added in a similar way is the $f$-Reggeon exchange process
  ($\mathbb{RRP}$ contribution, with $\mathbb{R}=f$),
   with $\alpha_f(0)=0.7$ \cite{kai09}
  and with  the corresponding probabilities  $w^{f}_h$.

\section{Results for secondary particle production\label{results.sec}}
 Regarding applications of MC event generators to EAS
simulations, of primary importance is an accurate enough description
of forward hadron production. 
Here comes a significance of experimental
measurements performed at fixed target energies, where forward spectra
of secondary particles have been studied in great detail.

\begin{figure*}[p]
\centering
\includegraphics[height=5.1cm,width=0.9\textwidth]{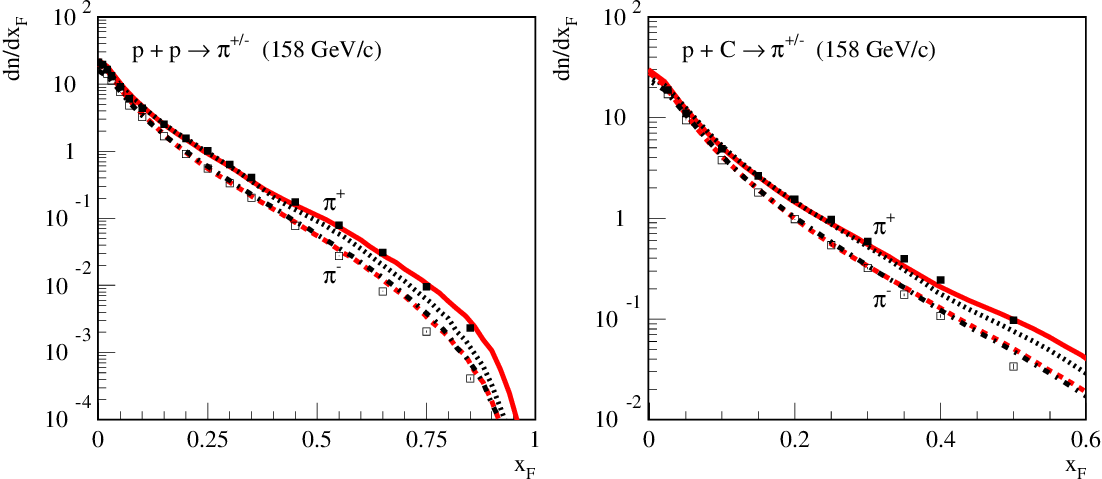}
\caption{$x_{\rm F}$-distributions of charged pions (in c.m.\ frame) in
$pp$ (left) and $p$C (right) collisions at 158 GeV/c, as calculated using the 
QGSJET-III  model, compared to NA49 data \cite{alt06,alt07} (points):
 $\pi^+$ -- solid lines and filled squares,  $\pi^-$ -- dashed lines and 
 open squares. The corresponding
results of the QGSJET-II-04 model are shown by dotted and dash-dotted
 lines for $\pi^+$ and  $\pi^-$, respectively.}
\label{fig:pion158}       
\end{figure*}

\begin{figure*}[p]
\centering
\includegraphics[height=5.1cm,width=0.9\textwidth]{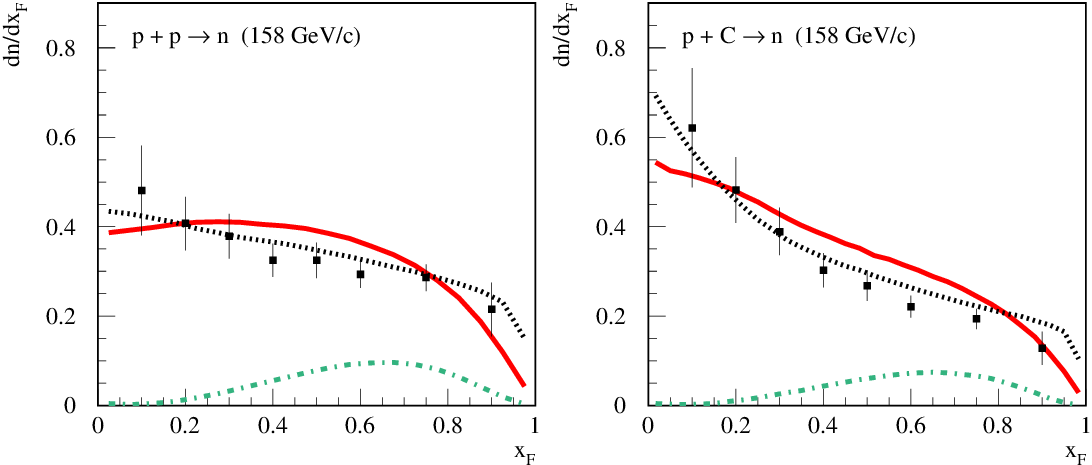}
\caption{$x_{\rm F}$-distributions of neutrons (in c.m.\ frame) in
$pp$ (left) and $p$C (right) collisions at 158 GeV/c, as calculated using the 
QGSJET-III  (solid lines) and QGSJET-II-04 (dotted lines) models, compared to NA49 data  \cite{ant10,baa13} (points). Partial contributions of the pion exchange process
in the  QGSJET-III model are shown by dash-dotted lines.}
\label{fig:neut158}       
\end{figure*}

\begin{figure*}[p]
\centering
\includegraphics[height=5.1cm,width=0.9\textwidth]{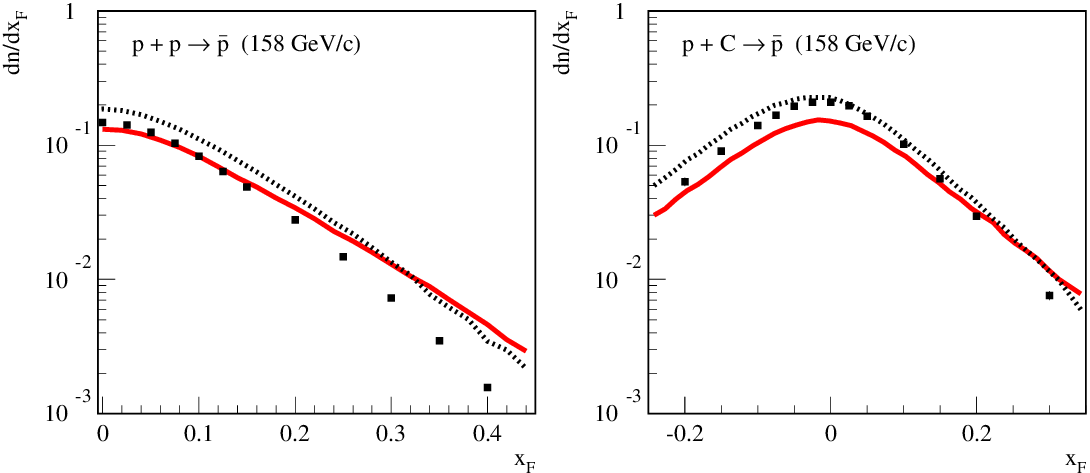}
\caption{$x_{\rm F}$-distributions of antiprotons (in c.m.\ frame) in
$pp$ (left) and $p$C (right) collisions at 158 GeV/c, as calculated using the 
QGSJET-III  (solid lines) and QGSJET-II-04 (dotted lines) models,
 compared to NA49 data \cite{ant10,baa13} (points).}
\label{fig:pbar158}       
\end{figure*}

We start in Fig.\ \ref{fig:prot158} 
 with invariant cross sections in c.m.\ frame for  proton production in $pp$ and $p$C
collisions at 158 GeV/c, comparing the corresponding results of the QGSJET-III model
to the ones of QGSJET-II-04 and to measurements by  the NA49 
experiment.
The same comparison regarding Feynman $x$ ($x_{\rm F}$) distributions of charged pions,
 neutrons, and antiprotons, also for  $pp$ and $p$C interactions at 158 GeV/c,
is presented in Figs.\ \ref{fig:pion158},  \ref{fig:neut158}, and  \ref{fig:pbar158}; 
 \begin{figure}[htb]
\centering
\includegraphics[height=6.cm,width=0.49\textwidth]{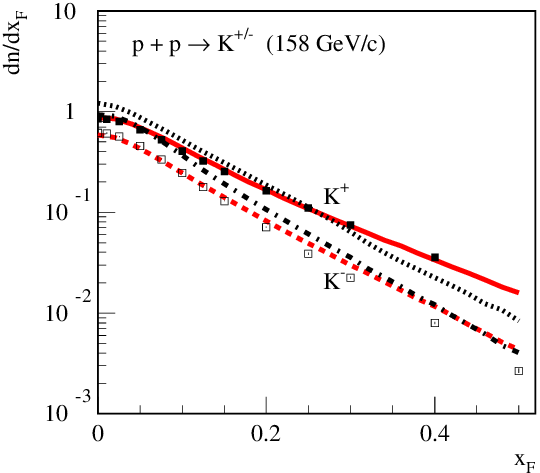}
\caption{$x_{\rm F}$-distributions of charged kaons (in c.m.\ frame) in
$pp$   collisions at 158 GeV/c, as calculated using the 
QGSJET-III  model, compared to NA49 data \cite{ant10a} (points): 
$K^+$ -- solid line and filled squares,  $K^-$ -- dashed line 
and open squares. The corresponding
results of the QGSJET-II-04 model are shown by dotted and dash-dotted
 lines for $K^+$ and  $K^-$, respectively.}
\label{fig:kaon158}       
\end{figure}%
the partial contribution to neutron production due to the  
pion exchange mechanism in the QGSJET-III model is 
shown separately in Fig.\  \ref{fig:neut158}. In turn, $x_{\rm F}$ distributions of
secondary charged kaons, for $pp$  collisions at 158 GeV/c,  calculated 
using  the  QGSJET-III  and QGSJET-II-04 models, are compared to NA49 data in Fig.\ \ref{fig:kaon158}.
 Regarding the production of $p$, $n$, $\pi^{\pm}$, and  $K^{\pm}$, we observe relatively
 small differences between QGSJET-III  and QGSJET-II-04, both models being in a satisfactory
 agreement with the measurements. This is, however, not the case for
  secondary antiprotons: 
  the predicted  $x_{\rm F}$ distributions are substantially harder
   than observed by the  experiment, especially, in the case of the QGSJET-III 
  model. One possible way to improve the agreement with the data is
  to allow for a separate treatment of the hadronization of the ``remnant''
  of the incoming proton, which, however, requires to introduce
  additional adjustable parameters to a model  \cite{rie20}. Generally, the problem
  is related to an extrapolation of the high energy treatment towards low
  energies \cite{kac15}. Correspondingly, its ultimate solution requires a full
  scale treatment of the transition into the low energy regime, taking into
  consideration  Reggeon exchanges  and the Reggeon-Reggeon-Reggeon ($\mathbb{RRR}$) contributions.

\begin{figure*}[htb]
\centering
\includegraphics[height=6.cm,width=\textwidth]{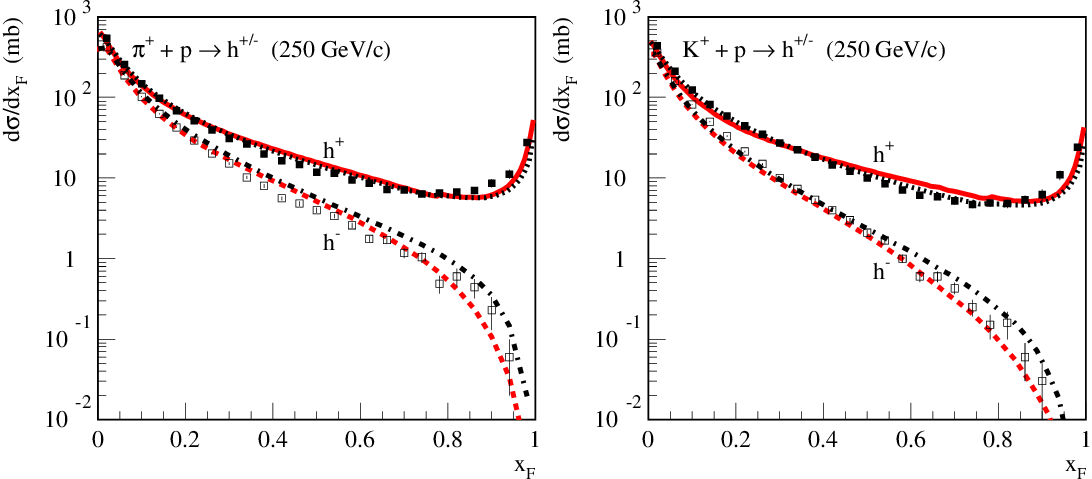}
\caption{$x_{\rm F}$-dependence of production cross sections in c.m.\ frame for charged hadrons ($h^{\pm}$) in $\pi^+p$ (left) and $K^+p$ (right)
 collisions at 250 GeV/c, as calculated using the  QGSJET-III  model, compared to NA22 data \cite{ada88} (points): 
$h^+$ -- solid lines and filled squares,  $h^-$ -- dashed lines and open squares. The corresponding results of the QGSJET-II-04 model are  
shown by dotted and dash-dotted lines for $h^+$ and  $h^-$, respectively.}
\label{fig:pipi250}       
\end{figure*}

\begin{figure*}[p]
\centering
\includegraphics[height=6.cm,width=\textwidth]{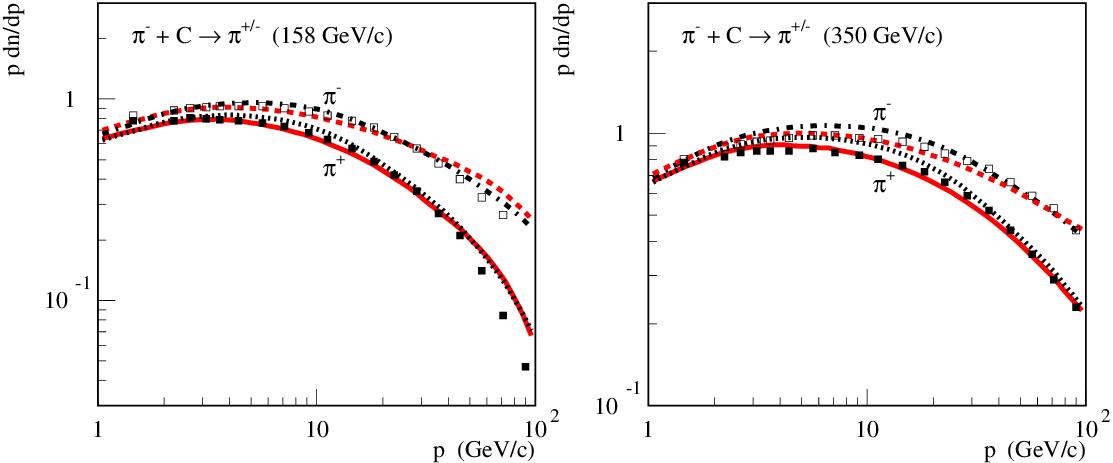}

\vspace{2mm}
\includegraphics[height=6.cm,width=\textwidth]{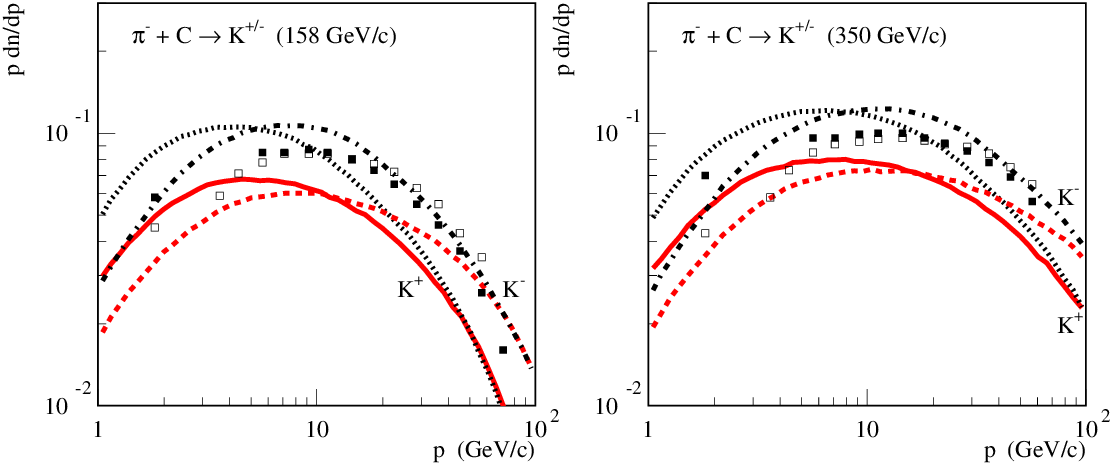}

\vspace{1mm}
\includegraphics[height=6.cm,width=\textwidth]{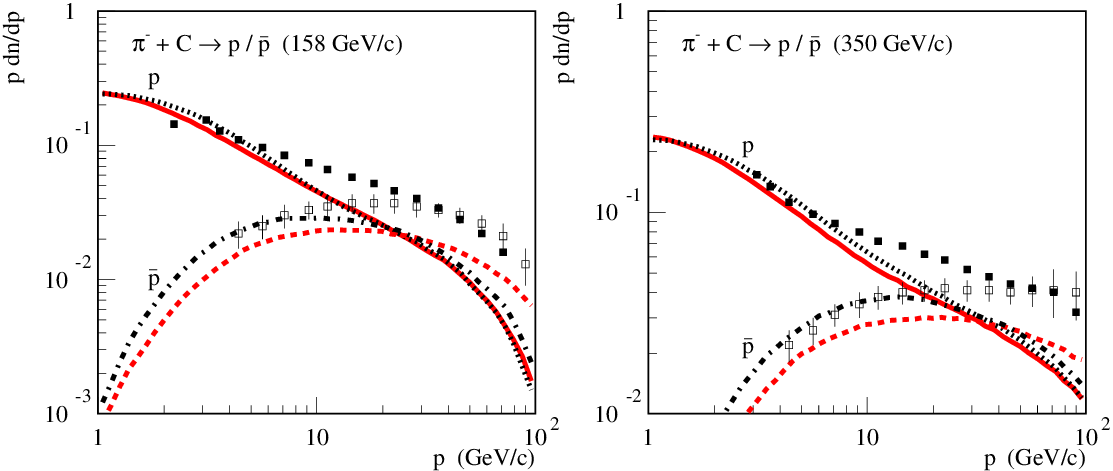}
\caption{Momentum distributions  in laboratory frame of charged pions (top panels),
  charged kaons (middle panels), protons and antiprotons (bottom panels), produced
  in $\pi^-C$ collisions at 158 GeV/c (left) and 350 GeV/c (right), 
  as calculated using the  QGSJET-III  model, compared to NA61 data \cite{adh23} (points).
  Solid lines and filled squares correspond to positively charged hadrons while 
   dashed lines and open squares refer to negatively charged hadrons.
  The corresponding results of the QGSJET-II-04 model are 
shown by dotted and dash-dotted lines for  positively and 
negatively charged hadrons, respectively.}
\label{fig:pimcpi}       
\end{figure*}

\begin{figure}[htb]
\centering
\includegraphics[height=6.cm,width=0.49\textwidth]{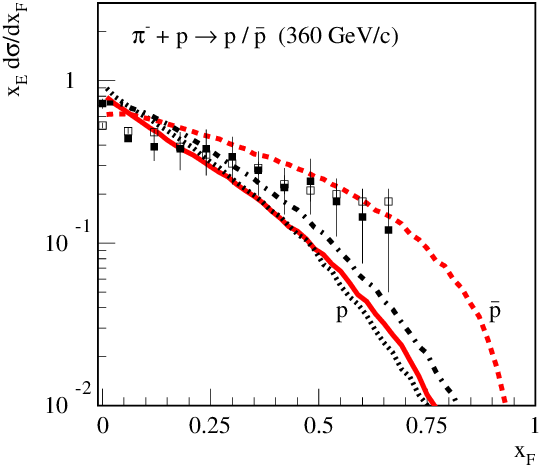}
\caption{$x_{\rm F}$-dependence of $p_t$-integrated invariant cross sections in 
c.m.\ frame, $x_E d\sigma /dx_{\rm F}$, for proton and antiproton production in
$\pi^-p$ collisions at  360 GeV/c, as calculated using the 
  QGSJET-III  model, compared to LEBC-EHS data \cite{agu87} (points): $p$ -- solid line and filled squares,  $\bar p$ -- dashed line and open squares. The corresponding
results of the QGSJET-II-04 model are shown by dotted and dash-dotted
 lines for $p$ and $\bar p$, respectively.}
\label{fig:pip360}       
\end{figure}

Further,  in Fig.\ \ref{fig:pipi250}, 
we compare $x_{\rm F}$ spectra of positively and negatively charged secondary
 hadrons, for $\pi^+p$ and $K^+p$ collisions at 250 GeV/c, calculated using the 
 QGSJET-III  and QGSJET-II-04   models, to measurements by the
  NA22 experiment \cite{ada88}. Additionally, in Fig.\ \ref{fig:pimcpi}, 
  we compare the   results of the two models for momentum
   distributions of various  secondary hadrons
 produced in $\pi^-C$ collisions at 158 and 350 GeV/c  to recent measurements
of the NA61 experiment \cite{adh23}. For both models, 
we observe an acceptable agreement with the data
both for the  charged particle spectra in Fig.\ \ref{fig:pipi250} and for the pion
distributions in  Fig.\ \ref{fig:pimcpi}.
However,  charged kaon production in $\pi^-C$ collisions 
appears to be underestimated by QGSJET-III
by $\sim 30$\%, compared to the NA61 data. 
This is somewhat surprising since no such
a ``kaon deficit''  was observed for $pp$ collisions (cf.\ Fig.\ \ref{fig:kaon158}). 
Even more striking is the considerable
underestimation  by the  QGSJET-III and QGSJET-II-04 models
 of   proton and antiproton production in $\pi^-C$ collisions,
  compared to the NA61 measurements. While these discrepancies may,
in principle,  be attributed to  potential deficiencies of the 
hadronization procedures of the models, it is noteworthy that no such a striking
underestimation of secondary proton and antiproton yields is observed if we
compare the predictions of  QGSJET-III for $p$ and $\bar p$ production in $\pi ^-p$
collisions at 360 GeV/c to the
data of the LEBC-EHS experiment \cite{agu87},
 as one can see in Fig.\  \ref{fig:pip360}.

\begin{figure}[htb]
\centering
\includegraphics[height=6.cm,width=0.49\textwidth]{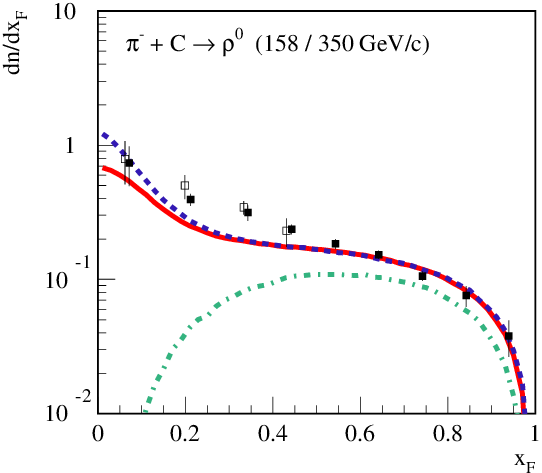}
\caption{$x_{\rm F}$-distributions of $\rho^0$-mesons (in c.m.\ frame) in
$\pi^-C$ collisions at 158 GeV/c (solid line) and 350 GeV/c (dashed line),
 as calculated using the  QGSJET-III  model, compared to NA61
  data \cite{adu17} (points). Partial contribution of the pion exchange process 
for $\pi^-C$ collisions at 158 GeV/c is  shown by a dash-dotted line.}
\label{fig:pimcrho}       
\end{figure}

\begin{figure*}[t]
\centering
\includegraphics[height=6.cm,width=\textwidth]{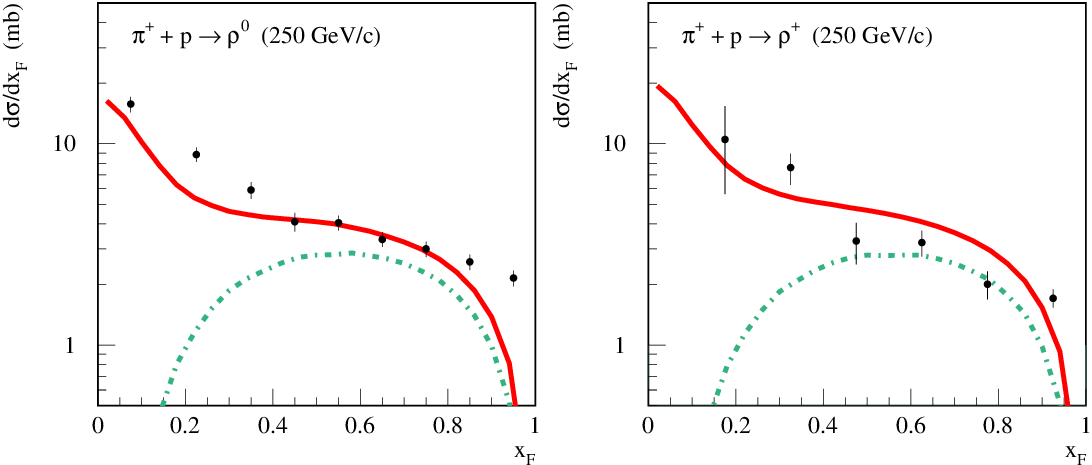}
\caption{$x_{\rm F}$-dependence of production cross sections in c.m.\ 
frame for $\rho^0$ (left) and $\rho^+$ (right) mesons in $\pi^+p$ 
 collisions at 250 GeV/c, 
as calculated using the  QGSJET-III  model, compared to NA22
 data \cite{aga90} (points).  Partial contributions of the pion 
 exchange process  are shown by dash-dotted lines.}
\label{fig:pirho250}       
\end{figure*}

As discussed in \cite{ost13}, of particular importance for model predictions
regarding EAS muon content is a satisfactory description of forward production
of $\rho$-mesons in pion-air collisions, which is dominated by the pion 
exchange process. In Fig.\  \ref{fig:pimcrho},
we compare  $x_{\rm F}$ distributions of  $\rho^0$ mesons produced in 
  $\pi ^-C$ collisions at 158 and 350 GeV/c, calculated using the  QGSJET-III  model, 
  to the respective data of the NA61 experiment,
 while Fig.\ \ref{fig:pirho250} contains a comparison of the model predictions
 for the spectra of  $\rho^0$ and  $\rho^+$ mesons produced in $\pi ^+p$
interactions at 250 GeV/c to measurements of the NA22 experiment. Both 
in Fig.\  \ref{fig:pimcrho} and in Fig.\ \ref{fig:pirho250}, 
we show also the partial contribution of the pion exchange mechanism
 to  $\rho$ meson  production, which clearly dominates the forward parts
 of the spectra. Overall, the agreement is quite
good, notably, when comparing to the recent measurements of NA61. 
 
\begin{figure}[htb]
\centering
\includegraphics[height=6.cm,width=0.49\textwidth]{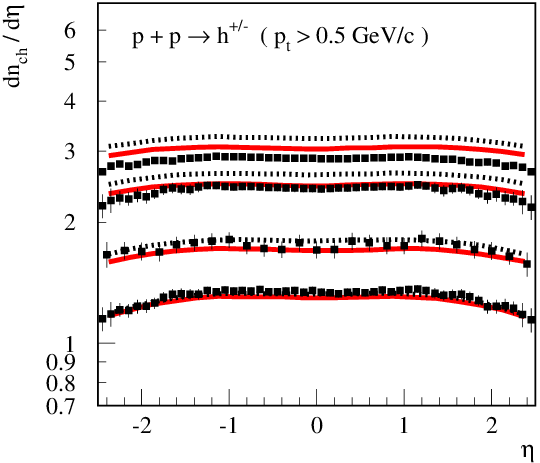}
\caption{Pseudorapidity distributions of charged hadrons (with $p_t>0.5$ GeV/c) in c.m.\ frame, 
produced in $pp$ collisions at different $\sqrt{s}$ (from top to bottom:
13, 7, 2.36, and 0.9 TeV), as calculated using the 
QGSJET-III  (solid lines) and QGSJET-II-04 (dotted lines) models, 
compared to ATLAS data \cite{aad11,aad16} (points).}
\label{fig:ppeta}       
\end{figure}

\begin{figure}[htb]
\centering
\includegraphics[height=6.cm,width=0.49\textwidth]{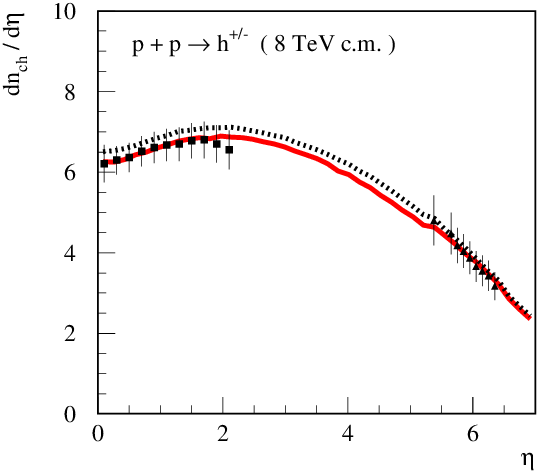}
\caption{Pseudorapidity distribution of charged hadrons  in c.m.\ frame, 
produced in $pp$ collisions at $\sqrt{s}=8$ TeV, as calculated using the 
QGSJET-III  (solid line) and QGSJET-II-04 (dotted line) models, 
compared to a simultaneous measurement by CMS and TOTEM \cite{cha14} (points).}
\label{fig:cmstot}       
\end{figure}

  Coming now to measurements at the Large Hadron Collider (LHC),
   those mostly provided information
  on hadron production in $pp$, $pA$, and $AA$ collisions, at 
  central ($y\sim 0$) rapidities in c.m.\ frame. A comparison of the
  pseudorapidity density of charged hadrons, $dN^{\rm ch}_{pp}/d\eta$,
  in $pp$ collisions at $\sqrt{s}=0.9$, 2.36, 7, and 13 TeV, calculated
  using the QGSJET-III  and QGSJET-II-04  models,  to the 
  respective ATLAS data\footnote{The experimental event selection requires at least
  one charged secondary hadron of $p_{t}>0.5$ GeV at
   $|\eta|<2.5$.} \cite{aad11,aad16}
    is presented in Fig.\  \ref{fig:ppeta}.
  Since these experimental data have been used to tune the QGSJET-III model,
     the overall satisfactory agreement comes at no surprise.
 Much less trivial is the agreement of the predictions of both models
  to the results of a  combined measurement\footnote{The experimental event 
 selection requires at least
  one charged secondary hadron  at $5.3<|\eta|<6.5$.}  by the  CMS and TOTEM 
  experiments of $dN^{\rm ch}_{pp}/d\eta$  over a wide $\eta$ range,
 in $pp$ collisions  at  $\sqrt{s}=8$ TeV \cite{cha14},
    see Fig.\  \ref{fig:cmstot}.
As discussed in \cite{ost16}, this set of experimental data allows one
to test very basic model assumptions regarding the momentum distributions
of constituent partons.

\begin{figure}[htb]
\centering
\includegraphics[height=6.cm,width=0.49\textwidth]{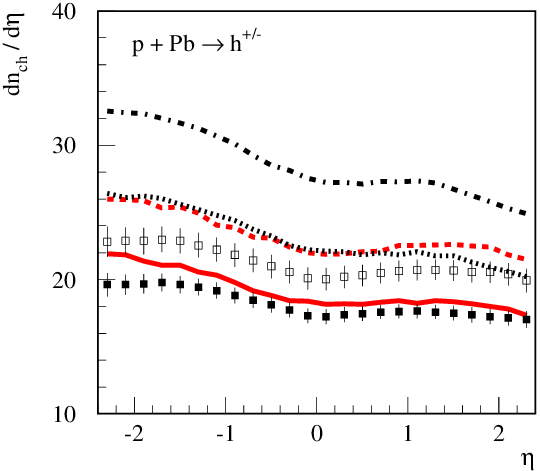}
\caption{Pseudorapidity distributions of charged hadrons  in c.m.\ frame, 
produced in $p$Pb collisions at different $\sqrt{s}$, as calculated using the 
QGSJET-III   model, compared to CMS data \cite{sir17} (points):
 $\sqrt{s}=5$ TeV --  solid line and filled squares,
  $\sqrt{s}=8$ TeV -- dashed line and open squares. The corresponding
results of the QGSJET-II-04 model are shown by dotted and dash-dotted
lines for  $\sqrt{s}=5$ TeV and  $\sqrt{s}=8$ TeV, respectively.}
\label{fig:ppb}       
\end{figure}

We also show in Fig.\  \ref{fig:ppb} 
 the calculated pseudorapidity density of charged hadrons, $dN^{\rm ch}_{p{\rm Pb}}/d\eta$,
for proton-lead collisions at   $\sqrt{s}=5$  and 8 TeV, compared to the data
of the CMS experiment \cite{sir17}. Here the results of the QGSJET-III model
agree with the measurements substantially better, compared to the ones of  QGSJET-II-04.

\begin{figure}[htb]
\centering
\includegraphics[height=6.cm,width=0.49\textwidth]{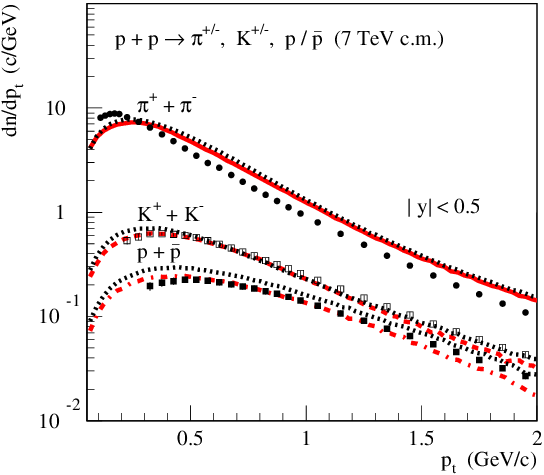}
\caption{Transverse momentum distributions of identified hadrons at central
rapidity ($|y|<0.5$)  in c.m.\ frame, for $pp$ collisions at  $\sqrt{s}=7$ TeV,
 calculated using the QGSJET-III   model, in comparison  to ALICE data \cite{ada15} (points): $\pi^+ + \pi^-$ --  solid line and filled circles,
  $K^+ + K^-$ -- dashed line and open squares,
    $p + \bar p$ -- dash-dotted line and filled squares. The corresponding
results of the QGSJET-II-04 model are shown by dotted lines.}
\label{fig:pt-spectra}       
\end{figure}

As an example comparison with identified secondary hadron spectra, we plot
in  Fig.\  \ref{fig:pt-spectra} transverse momentum distributions of charged
pions,  charged kaons, and protons plus antiprotons, produced in $pp$ collisions at   $\sqrt{s}=7$ TeV, calculated with the  QGSJET-III and QGSJET-II-04 models,
being confronted to the corresponding data of the ALICE experiment.
The somewhat deficient description of these  distributions by the two models
may be related to the simplified string fragmentation procedure employed,
based on the duality principle: not treating explicitly the production and 
decays of short-lived hadronic resonances. 

\begin{table*}[t]
\begin{tabular*}{1\textwidth}{@{\extracolsep{\fill}}llllll}
\hline 
$M_{X}$ range, GeV & $3.1-7.7$  &  $7.7-380$  & $380-1150$  &  $1150-3100$ &
  $3.1-3100$\tabularnewline
\hline 
\hline 
TOTEM \cite{olj20} & $1.83\pm 0.35$ & $4.33 \pm 0.61$ & $2.10 \pm 0.49$
 & $2.84 \pm 0.40$ & $11.10 \pm 1.66$ \tabularnewline
QGSJET-II-04 & 1.88 (0) & 3.21 (0.07) & 1.06 (0.57) & 5.91 (5.53) & 
12.06 (6.17)\tabularnewline
QGSJET-III & 1.41 (0.04) & 3.19 (0.24) & 1.51 (0.44) & 6.38 (4.97) & 
12.49 (5.69) \tabularnewline
\hline 
\end{tabular*}
\caption{Predictions of the QGSJET-II-04 and QGSJET-III models for
cross sections of SD-like events (in mb) at $\sqrt{s}=7$ TeV, for different
ranges of mass $M_{X}$ of diffractive states produced, 
compared to TOTEM data  \cite{olj20}. Values in brackets correspond to the
respective contributions of nondiffractive events.
\label{tab: SD-totem}}
\end{table*}
However, regarding the relevance to EAS calculations, of primary significance
is the energy-dependence of forward hadron production. In that respect,
of particular importance are diffraction studies at LHC.
 In Table \ref{tab: SD-totem}, we
compare the predictions of the  QGSJET-III and   QGSJET-II-04 models,
 regarding partial cross
sections for single diffraction-like (SD-like) events
at  $\sqrt{s}=7$ TeV, for different intervals of the diffractive 
state mass $M_X$, to the respective data of the TOTEM experiment \cite{olj20}.
While the cumulative cross section for the full range $3.1 <M_X<3100$ GeV,
studied by TOTEM, is well in agreement between both models and the data,
the predicted rates of  SD-like events for intermediate diffractive masses,
 $7.7 <M_X<1150$ GeV, are somewhat lower than observed by TOTEM.
On the other hand, the production of large mass states,  $1150 <M_X<3100$ GeV,
is overestimated in both models by more than a factor of two, compared to the
TOTEM measurements.  Taking into account that the event rate in this mass
interval is dominated by a formation of random rapidity gaps in
 nondiffractive hadron production (see the corresponding values in 
brackets in Table  \ref{tab: SD-totem}), primarily,  by  $\mathbb{RRP}$ processes
 (see, e.g.\ \cite{kmr10} for the corresponding discussion), this may
  indicate a certain overestimation of the  $\mathbb{RRP}$ contributions
   or/and some deficiencies of the hadronization procedures of both models.

\begin{figure*}[p]
\centering
\includegraphics[height=11.7cm,width=\textwidth]{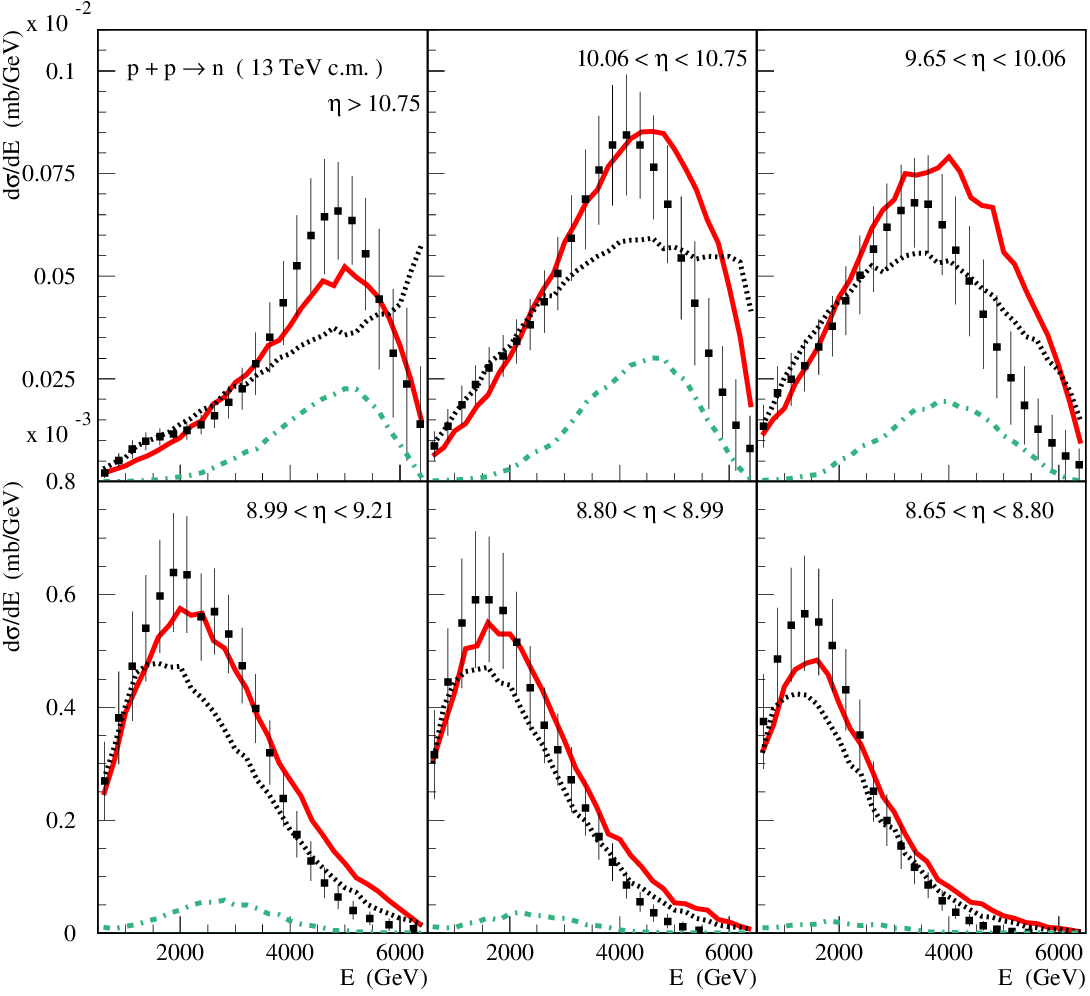}
\caption{Neutron energy spectra (in c.m.\ frame)
 in  $pp$ collisions at $\sqrt{s}=13$ TeV,   
calculated using the QGSJET-III (solid lines) and  QGSJET-II-04 
(dotted lines) models, for different pseudorapidity intervals 
 (as indicated
in the plots), in comparison to LHCf data \cite{adr18} (points).
 Partial contributions of the pion exchange process in the
 QGSJET-III  model   are shown by dash-dotted lines.}
\label{fig:nlhcf13}       
\end{figure*}

\begin{figure*}[p]
\centering
\includegraphics[height=7.cm,width=\textwidth]{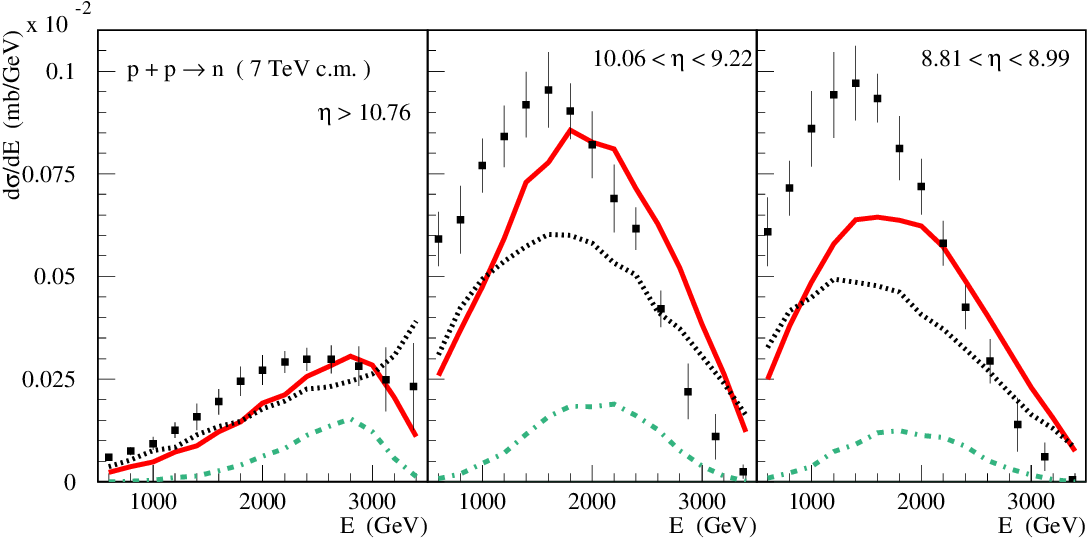}
\caption{Same as in Fig.\ \ref{fig:nlhcf13}, for  $\sqrt{s}=7$ TeV.}
\label{fig:nlhcf7}       
\end{figure*}

Regarding the so-called inelasticity, i.e., the relative energy loss of
most energetic secondary hadrons, important constraints come from 
measurements of forward neutron production by the LHCf 
experiment \cite{adr15,adr18}. When interpreting those experimental results, it is
important to take into consideration the contribution of the $t$-channel pion
exchange \cite{kai06,kop15,kmr17}. Here the  data of the NA49 experiment  on
neutron production in $pp$ and $p$C collisions at 158 GeV/c \cite{ant10,baa13}
(cf.\ Fig.\ \ref{fig:neut158}) can be combined with
those of LHCf at LHC energies in order to test the implementation of the pion
exchange mechanism in  the QGSJET-III model \cite{ost21}, notably,
 regarding the predicted energy-dependence over
a wide c.m.\ energy range,  $17<\sqrt{s}<13000$ GeV.
 Comparing the calculated neutron energy
spectra to the observations in Figs.\   \ref{fig:nlhcf13}-\ref{fig:nlhcf7}, 
we see a satisfactory agreement of the model with the measurement at $\sqrt{s}=13$ TeV,
while there is a certain underestimation of the neutron yield at 
  $\sqrt{s}=7$ TeV, compared to LHCf data.\footnote{As one can see in Figs.\
  \ref{fig:neut158} and  \ref{fig:nlhcf13}, the agreement with the data on
  neutron production, both from the NA49 experiment at 158 GeV/c and from
  LHCf at $\sqrt{s}=13$ TeV, can be improved considering a smaller
  contribution of the pion exchange process,
   i.e., choosing a smaller value for 
  the $w^{\pi}_p$ parameter. This would, however, aggravate the tension
  with the LHCf data at  $\sqrt{s}=7$ TeV.}
   The latter is actually  surprising, 
  given the small difference between the two c.m.\ energies for LHCf studies.

\begin{figure*}[t]
\centering
\includegraphics[height=6cm,width=\textwidth]{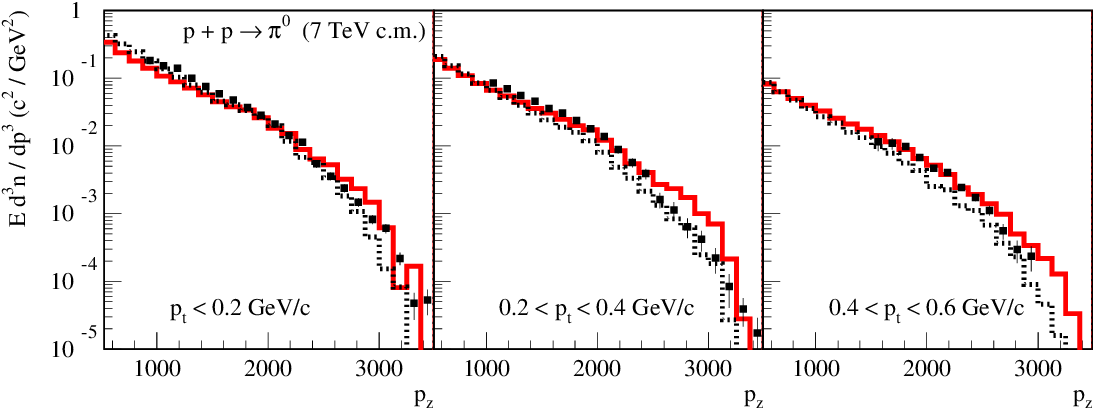}
\caption{Invariant momentum spectra of neutral pions in  c.m.~frame,
$E\, d^3n_{pp}^{\pi^{0}}/dp^3$, for different $p_t$ selections (as indicated
in the plots),  for $pp$ collisions at $\sqrt{s}=7$ TeV, as calculated 
 using the QGSJET-III (solid histograms) and  QGSJET-II-04 
(dotted histograms) models,
 compared to the data  of the LHCf experiment \cite{adr16} (points).}
\label{fig:lhcfpi}      
\end{figure*}

Another important benchmark from the LHCf experiment concerns the measured forward spectra
of neutral pions \cite{adr16}, which is also of importance for astrophysical
studies based on $\gamma$-rays (see, e.g.\ \cite{kmo19,kol21} for the corresponding
discussion). The results of the QGSJET-III and  QGSJET-II-04 models 
are confronted to the LHCf data in Fig.\ \ref{fig:lhcfpi}.
While both models demonstrate a satisfactory agreement with  the measurements,
somewhat disturbing is a certain underestimation of $\pi^0$ production at
$p_z\sim 1$ TeV by the QGSJET-III model, which may be related to its softer
momentum distribution  of constituent partons [larger $\alpha_{\rm sea}$, 
cf.\ Eq.\ (\ref{ems.eq})].

\section{Basic characteristics of proton-nitrogen collisions\label{pn.sec}}
Before proceeding  with the model application to extensive air shower modeling,
it is worth considering its predictions for some basic characteristics of
hadron-nucleus collisions, relevant for EAS predictions.

\begin{figure}[htb]
\centering
\includegraphics[height=6.cm,width=0.49\textwidth]{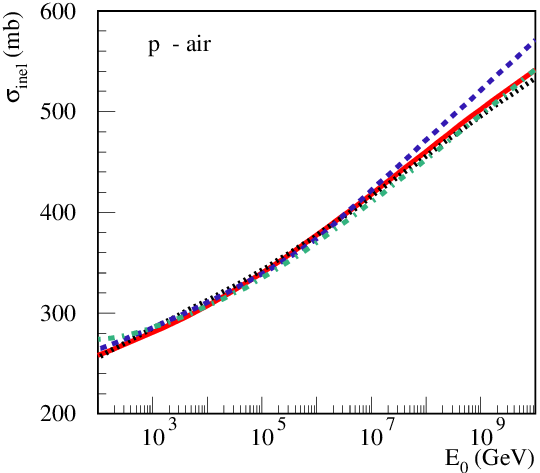}
\caption{Energy-dependence of   $\sigma^{\rm inel}_{p-{\rm air}}$,  calculated 
using the QGSJET-III, QGSJET-II-04, EPOS-LHC, and SIBYLL-2.3 models --
solid, dotted, dashed, and dash-dotted lines, respectively. }
\label{fig:sigpair}       
\end{figure}

The first quantity to consider is the  inelastic proton-air
cross section $\sigma^{\rm inel}_{p-{\rm air}}$ which defines the mean free path
(m.f.p.) $\lambda_p = m_{\rm air}/\sigma^{\rm inel}_{p-{\rm air}}$ 
($m_{\rm air}$ being the average mass of air nuclei)
of a primary CR proton in the atmosphere and impacts thereby many
EAS characteristics, notably, the EAS maximum depth $X_{\max}$
corresponding to the position of the maximum of charged particle profile of an
air shower. The energy-dependence of  $\sigma^{\rm inel}_{p-{\rm air}}$,
  calculated using the
QGSJET-III, QGSJET-II-04, EPOS-LHC \cite{pie15}, and SIBYLL-2.3 \cite{rie20}
 models, is plotted in Fig.\ \ref{fig:sigpair}.  
 Not surprisingly, there is a good agreement between the predictions of 
QGSJET-III and QGSJET-II-04, as well as with the ones of the other two models
considered, since more or less the same set of accelerator data on the
total and elastic proton-proton cross sections had been used for the
relevant parameter tuning in all the cases.  The small differences between
the results for  $\sigma^{\rm inel}_{p-{\rm air}}$, 
all obtained using the Glauber-Gribov
formalism \cite{gla56,gri69}, may stem from a choice of the nuclear ground state
wave function or/and from a model treatment of the inelastic screening effects.
With the most steep energy rise of    $\sigma^{\rm inel}_{p-{\rm air}}$ 
being predicted
by  EPOS-LHC, its cross section is $\simeq 5$\% higher at $E_0=10^{10}$ GeV than the
one of QGSJET-III. Yet the potential impact of that difference on 
$X_{\max}$ at the highest energies is $<3$ g/cm$^2$. 

\begin{figure}[htb]
\centering
\includegraphics[height=6.cm,width=0.49\textwidth]{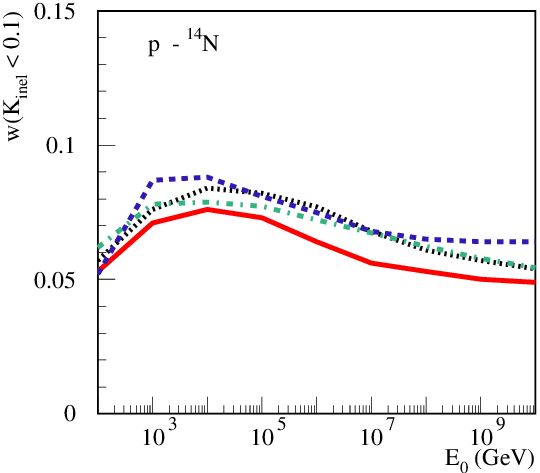}
\caption{Energy-dependence of  the probability of  diffractive-like interactions,
for  $p\,^{14}$N collisions, calculated using  different models.
The notations for the lines are the same as in Fig.\   \ref{fig:sigpair}.}
\label{fig:wdiffr}       
\end{figure}

Next, we consider the rate of the inelastic diffraction, which may also impact
 $X_{\max}$ predictions noticeably because of its typically small inelasticity
$K_{\rm inel}$,  i.e., the relative energy loss of leading (most energetic)
secondary nucleons. 
 This is particularly so for target diffraction characterized at very high
 energies by a tiny
 inelasticity $K_{\rm inel}\simeq M_X^2/s$, $M_X$ being the diffractive state
 mass. The effect of such ``quasi-elastic'' collisions is equivalent to a
 reduction of the inelastic proton-air cross section, giving rise to an
 increase of m.f.p\  of  primary CR protons:
 \begin{equation}
 \lambda_p \rightarrow \lambda_p(1+\sigma^{\rm diffr}_{p-{\rm air}}/\sigma^{\rm inel}_{p-{\rm air}})\,.
 \label{lambda-p.eq}
\end{equation}
In Fig.\  \ref{fig:wdiffr}, 
we compare the energy-dependence of the rate of diffractive-like interactions,
for proton-nitrogen collisions, i.e., the fraction of events characterized by
$K_{\rm inel}<0.1$, predicted by 
different models. It is worth reminding that, apart from diffractive interactions,  a sizable contribution to this rate comes from a formation of
random rapidity gaps in nondiffractive  collisions, 
notably, by  $\mathbb{RRP}$ processes. While the results of QGSJET-III and QGSJET-II-04 agree with each other within 10\%, the overall spread of the model
predictions reaches here 30\%, stemming both from different approaches to
the treatment of diffraction and from differences regarding the hadronization
procedure,  the latter potentially having a strong impact on the probability
to create random rapidity gaps. Yet, since such a variation of   the rate of diffractive-like interactions can produce only $\simeq 3$\% change of $\lambda_p$,
the corresponding impact  on  $X_{\max}$  is limited by $\simeq 2$ g/cm$^2$ 
[cf.\ Eq.\ (\ref{lambda-p.eq})].

\begin{figure}[htb]
\centering
\includegraphics[height=6.cm,width=0.49\textwidth]{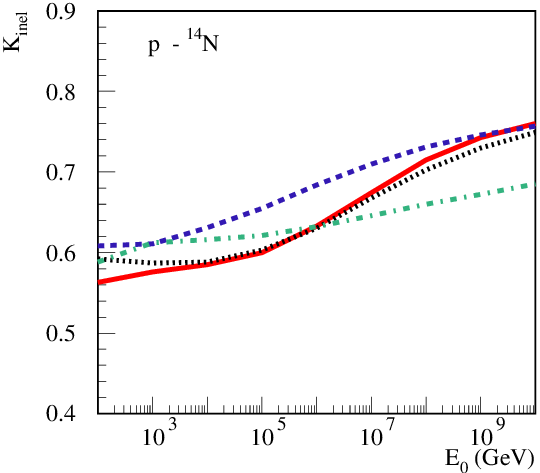}
\caption{Energy-dependence of  the inelasticity of $p\,^{14}$N collisions, 
calculated using   different models.
The notations for the lines are the same as in Fig.\   \ref{fig:sigpair}.}
\label{fig:kinel}       
\end{figure}

  Further we consider the inelasticity $K_{\rm inel}$ 
  of general proton-nitrogen collisions,
  which impacts strongly  the energy loss of leading secondary nucleons in the
  cause of the nuclear cascade development,  influencing significantly
   the predicted   $X_{\max}$ (see, e.g.\ \cite{ulr11}). The  energy
  dependence of  $K_{\rm inel}$, predicted by different models, is plotted 
  in    Fig.\ \ref{fig:kinel}.
Despite large differences between QGSJET-II-04 and QGSJET-III, both  regarding 
 the new developments in the latter, notably, the treatment of higher twist
 corrections to hard parton processes \cite{ost24,ost19} and the implementation
 of the pion exchange process \cite{ost21},  and concerning the values of important
 model parameters (e.g., the parameter $\alpha_{\rm sea}$ governing the momentum
 distribution of constituent partons), their predictions for  $K_{\rm inel}$
 are remarkably identical, both for the energy dependence and for the absolute
 values. A qualitatively similar behavior of  $K_{\rm inel}$ is predicted by
  EPOS-LHC. In contrast, in case of the SYBYLL-2.3 model, the inelasticity
  depends rather weakly on the collision energy, which is related to very basic
  model assumptions regarding the structure of constituent parton Fock 
  states and  impacts strongly the model results for  $X_{\max}$ \cite{ost16}.

\begin{figure}[htb] 
\centering
\includegraphics[height=6.cm,width=0.49\textwidth]{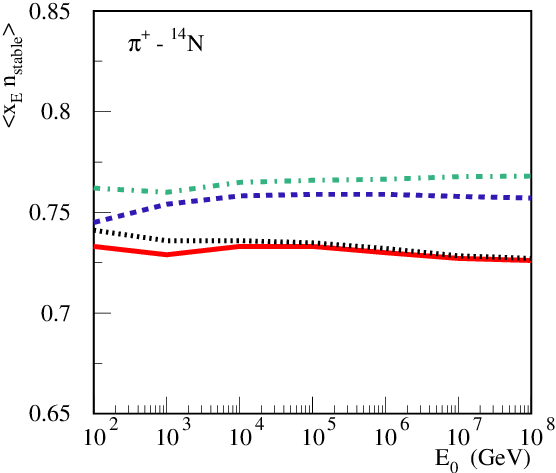}
\caption{Energy-dependence of  $\langle x_E\, n_{\rm stable}\rangle$,
 for $\pi^+\,^{14}$N collisions, calculated using  different models.
The notations for the lines are the same as in Fig.\   \ref{fig:sigpair}.}
\label{fig:nstable}       
\end{figure}

Regarding the EAS muon content,  the simple qualitative
model of Heitler \cite{hei54} implies a correlation of this observable with the
 number of ``stable'' secondary hadrons 
 [(anti)nucleons,  kaons, and charged pions], 
i.e., those which
have significant chances to interact in the atmosphere, before decaying.
 More precisely,  since the cascade nature of extensive 
  air showers enhances the importance
  of forward hadron production, the relevant quantity is the second
  moment of the distribution $dn_{\rm stable}/dx_E$ of 
 the energy fraction $x_E=E/E_0$ 
  of such hadrons 
  \cite{hil97}
   (see also \cite{rei21,ost24a} for  recent studies),  i.e., the total fraction
   of the parent hadron energy taken by all stable hadrons,
    \begin{equation}
 \langle x_E\, n_{\rm stable}(E_0)\rangle = \int_0^1 \!dx_E\,x_E\,\frac{dn_{\rm stable}(E_0,x_E)}{dx_E}\,.
 \label{xe.eq}
\end{equation}
 In  Fig.\  \ref{fig:nstable}, we compare  the energy-dependence of this quantity 
 for pion-nitrogen collisions,   calculated using the QGSJET-III, 
 QGSJET-II-04, EPOS-LHC, and SIBYLL-2.3 models. While there are only
 minor differences between  QGSJET-III and QGSJET-II-04 for the predicted
 $\langle x_E\, n_{\rm stable}\rangle$, the results of the other two
 models are quite different, which is related to a copious forward
 production of, respectively, (anti)baryons and $\rho$-mesons in the EPOS-LHC
 and SIBYLL-2.3 models.\footnote{See, however, \cite{ost23} for the corresponding
 criticism.}
   
  Overall, the results of the QGSJET-III and QGSJET-II-04 models, plotted in
   Figs.\ \ref{fig:sigpair}--\ref{fig:nstable}, appear to be rather similar
    to each other, suggesting that
   EAS predictions of the two models should not differ significantly.
  
\section{Predictions for EAS characteristics\label{eas.sec}}
\begin{figure*}[t]
\centering
\includegraphics[height=6.cm,width=\textwidth]{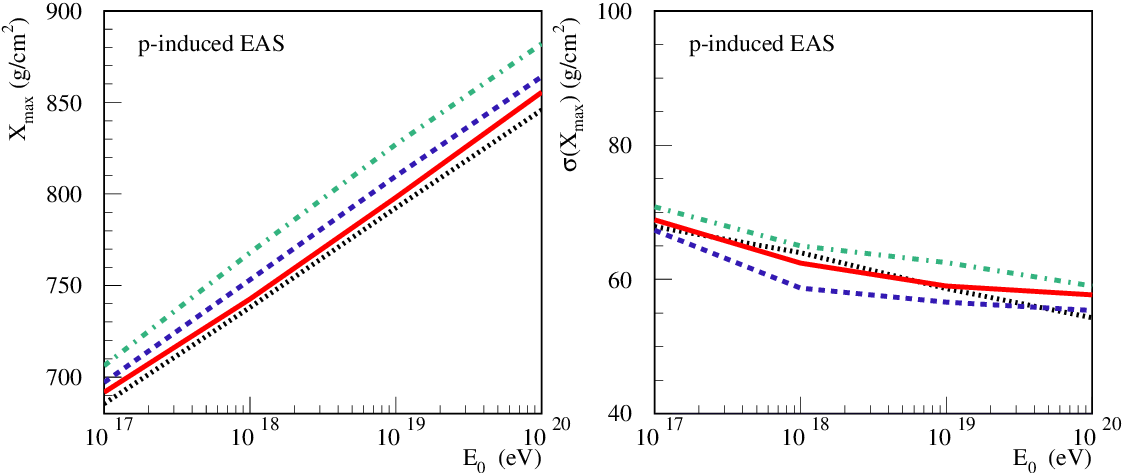}
\caption{Dependence on primary energy of  the maximum depth $X_{\max}$ (left)
and of its fluctuations  $\sigma(X_{\max})$ (right)
of proton-initiated EAS,
calculated using the 
QGSJET-III, QGSJET-II-04, EPOS-LHC, and SIBYLL-2.3 models --
solid, dotted, dashed, and dash-dotted lines, respectively.}
\label{fig:xmax}       
\end{figure*}

In Fig.\ \ref{fig:xmax}, 
 predictions of the QGSJET-III,  QGSJET-II-04, EPOS-LHC, and SIBYLL-2.3 models
  for the  energy-dependence of the maximum depth
$X_{\max}$ and of its fluctuation, $\sigma_{X_{\max}}$, of proton-induced 
extensive air
showers are compared. As anticipated in Section \ref{pn.sec}, the results of 
QGSJET-III and QGSJET-II-04 are rather similar to  each other, their 
$X_{\max}$ values differing by less than 10 g/cm$^2$.
Such an agreement is very impressive, given the substantial differences with
the predictions of the other two models.
 In turn, there is a very good
agreement between all the models, regarding the calculated values of $\sigma_{X_{\max}}$. This is not surprising since the quantity is mainly
defined by the inelastic proton-air cross section (see, e.g.\ \cite{kam12}),
while the impact of uncertainties regarding the rate of inelastic diffraction
does not exceed few  g/cm$^2$ \cite{ost14}.

\begin{figure}[htb]
\centering
\includegraphics[height=6.cm,width=0.49\textwidth]{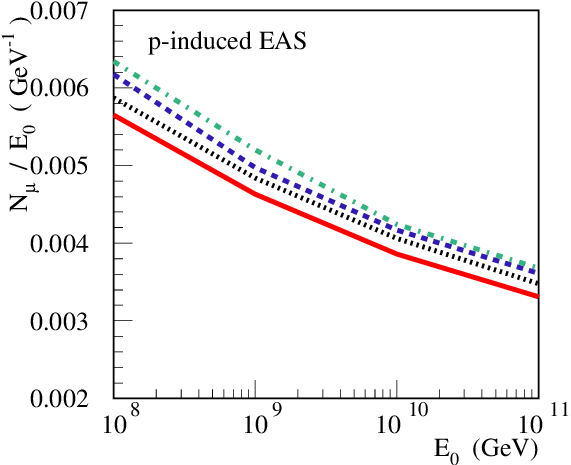}
\caption{Dependence on primary energy of the muon number  $N_{\mu}$
 (at sea level) of proton-initiated EAS, for $E_{\mu}>1$ GeV,
  calculated using   different models.
 The meaning of the  lines is the same as in  Fig.\ \ref{fig:xmax}.}
\label{fig:nmu}       
\end{figure}

Further, the calculated  energy-dependence of the muon number $N_{\mu}$ 
 (at sea level) of proton-initiated EAS, for muon energies $E_{\mu}>1$ GeV,
  is compared between the   models in  Fig.\  \ref{fig:nmu}. 
Here again,  the results of QGSJET-III and QGSJET-II-04 agree with each
other within 5\%. On the other hand, higher   $N_{\mu}$ 
predicted by EPOS-LHC and SIBYLL-2.3 comes at no surprise, given their
larger values for  $\langle x_E\, n_{\rm stable}\rangle$ 
(cf.\ Fig.\ \ref{fig:nstable} and the corresponding discussion
 in Section \ref{pn.sec}), being due to an enhanced 
forward production of, respectively, (anti)baryons and $\rho$-mesons in
those models.

\begin{figure}[htb]
\centering
\includegraphics[height=6.cm,width=0.49\textwidth]{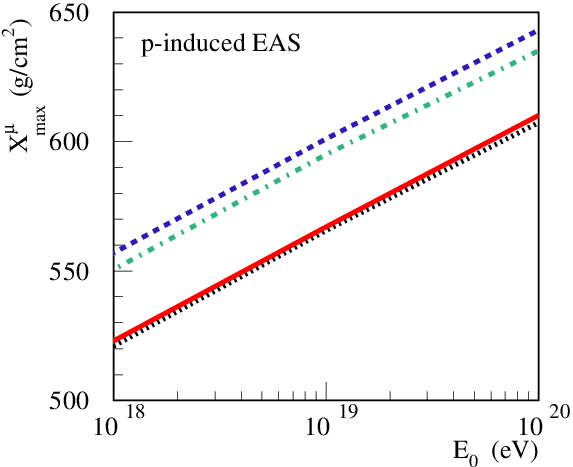}
\caption{Dependence on primary energy of the  maximal muon
production depth $X^{\mu}_{\max}$ ($E_{\mu}>1$ GeV)
 of proton-initiated 
 EAS,  calculated using  different models.
 The meaning of the  lines is the same as in  Fig.\ \ref{fig:xmax}.}
\label{fig:xmumax}       
\end{figure}

Finally, plotted in Fig.\ \ref{fig:xmumax} is 
 the   energy-dependence of the maximal muon production
  depth $X^{\mu}_{\max}$, i.e., the depth in the atmosphere corresponding to a maximal number of muons being produced by hadron decays, 
 calculated using the different models. As discussed in \cite{ost16a},
  this quantity is very sensitive to various aspects
of pion-air collisions and, thus, can be used to test and constrain the
respective model treatment, based on CR data. Even here, 
despite the fact that the energy dependence of hadron production in
pion-nucleus collisions is weakly constrained by available
 accelerator data, the agreement
between the results of QGSJET-III and QGSJET-II-04 is surprisingly good. 
On the other hand,
the substantially larger values of  $X^{\mu}_{\max}$, predicted by
the  EPOS-LHC and SIBYLL-2.3 models,
appear to be in a strong contradiction to the corresponding
measurements by the Pierre Auger Observatory \cite{aab14}.

\section{Conclusions\label{concl.sec}}
In this work, we discussed in some detail the hadronization procedure of the
QGSJET-III MC generator and presented selected results of that generator,
 regarding secondary hadron production, in comparison to experimental
  data and to the 
corresponding results of the previous model version, QGSJET-II-04. Overall, the
quality of agreement with the data for  QGSJET-III remained at approximately
the same level, as in the case of QGSJET-II-04, certain improvements being
mostly related to the treatment of the pion exchange process in hadron-proton
and hadron-nucleus collisions.

Comparing the predictions of different MC generators, 
regarding basic  characteristics of proton-induced extensive air showers,
 we observed an outstanding agreement between the results of  QGSJET-III 
and QGSJET-II-04, in spite of an implementation
of new theoretical approaches 
and a number of technical improvements  in the new model.
This may indicate that the treatment of relevant aspects of hadronic
interaction physics is already sufficiently constrained by available
accelerator data. On the other hand, in view of rather different results of
the other MC generators, regarding, e.g.,
 the air shower  maximum depth or EAS muon
content, one may rise a question whether the similarity of 
 extensive air shower predictions of   QGSJET-III 
and QGSJET-II-04 is rather a  consequence of the underlying
theoretical approaches shared by the two models or/and common deficiencies 
regarding the treatment of certain aspects of the interaction physics.
 Therefore, a general analysis
of potential model uncertainties for EAS predictions is a warranted task.
Such an analysis will be presented elsewhere \cite{ost24a,ost24b}.

\subsection*{Acknowledgments}

This work was  supported by  Deutsche Forschungsgemeinschaft 
(project number 465275045).

\end{document}